# Laser spectroscopic characterization of the nuclear clock isomer $^{229m}$Th

Johannes Thielking[1], Maxim V. Okhapkin[1], Przemysław Głowacki[1,†],

David M. Meier[1], Lars von der Wense[2], Benedict Seiferle[2],

Christoph E. Düllmann[3,4,5], Peter G. Thirolf[2], Ekkehard Peik[1]

1 Physikalisch-Technische Bundesanstalt, 38116 Braunschweig, Germany.
2 Ludwig-Maximilians-Universität München, 85748 Garching, Germany.
3 GSI Helmholtzzentrum für Schwerionenforschung GmbH, 64291 Darmstadt, Germany.
4 Helmholtz-Institut Mainz, 55099 Mainz, Germany.
5 Johannes Gutenberg-Universität, 55099 Mainz, Germany.

**The isotope $^{229}$Th is the only nucleus known to possess an excited state $^{229m}$Th in the energy range of a few electron volts, a transition energy typical for electrons in the valence shell of atoms, but about four orders of magnitude lower than common nuclear excitation energies. A number of applications of this unique nuclear system, which is accessible by optical methods, have been proposed. Most promising among them appears a highly precise nuclear clock that outperforms existing atomic timekeepers. Here we present the laser spectroscopic investigation of the hyperfine structure of $^{229m}$Th$^{2+}$, yielding values of fundamental nuclear properties, namely the magnetic dipole and electric quadrupole moments as well as the nuclear charge radius. After the recent direct detection of this long-searched-for isomer, our results now provide detailed insight into its nuclear structure and present a method for its non-destructive optical detection.**

Investigating $^{229}$Th with its unique low-energy transition between the nuclear ground state and a long-lived isomer at about 7.8 eV excitation energy provides the possibility to apply the precision methods of laser spectroscopy to a nuclear transition[1–3]. Building up on this, the realization of a nuclear clock has been proposed, i.e., an optical nuclear clock that uses the low-energy transition as the frequency reference. It is expected to benefit from the smaller sensitivity of the nucleus to external perturbations, including frequency shifts from electromagnetic fields, compared to electronic transitions exploited in current atomic clocks[4,5]. This application of the nuclear transition requires experimental efforts to study the nuclear properties, excitation methods of the isomeric state as well as its non-destructive observation[6].

The nuclear structure of $^{229}$Th has been studied via γ-ray spectroscopy of the radiation emitted after the α decay of $^{233}$U (refs 7–9), after β decay of $^{229}$Ac (refs 10,11), in Coulomb excitation[12] and in a (d,t) transfer reaction with $^{230}$Th (ref. 13). The low-lying levels can be assigned to rotational bands described by the Nilsson model[14]. The $^{229}$Th nuclear ground state is categorized as the bandhead of a 5/2+[633] rotational band and its nuclear moments have been determined experimentally[12,15]. A second rotational band has been identified as 3/2+[631]. Its bandhead – the low-energy isomer $^{229m}$Th that is investigated here – is still unresolved in γ spectroscopy.

---

† present address: Poznan University of Technology, 60-965 Poznań, Poland.

The transition energy between both levels has been determined indirectly as a difference of γ energies of intraband and interband transitions, resulting in the value 7.8(5) eV (refs 16,17). This corresponds to ultraviolet radiation at the wavelength of 160(10) nm, where the present uncertainty is about 17 orders of magnitude bigger than the expected natural linewidth. Depending on the electronic structure, the isomer may decay quickly via internal conversion[2,3] or radiatively with an estimated natural (i.e. unperturbed) lifetime of a few 1000 s (refs 11,18–20). The isomer's nuclear moments were estimated from nuclear structure models[18,21,22]. Many experimental attempts to induce and detect an optical excitation of the isomer have failed so far, impeded by the difficulty to produce widely tunable intense vacuum-ultraviolet radiation, by background of ionizing radiation from the $^{229}$Th samples, and by competition with non-radiative relaxation processes[6,23,24]. Apart from the spectroscopic determination of the nuclear spin and indirect measurements of the excitation energy[7,8,16,17], no experimental data on the nuclear properties of the isomer were available until recently. In a novel approach, using recoil ions from the decay of $^{233}$U as a source of $^{229m}$Th, electrons emitted in the internal conversion decay of the isomer in neutral thorium were detected[25] and the half-life for this process was measured[26].

The availability of the isomer in recoil ions opens a way to measure the unknown nuclear properties of $^{229m}$Th via laser spectroscopy of electronic transitions. We report the optical detection of ions in the $^{229m}$Th isomeric state and the resolved hyperfine structure (HFS), which arises from the interaction of the isomer nucleus with the valence electrons (see Methods).

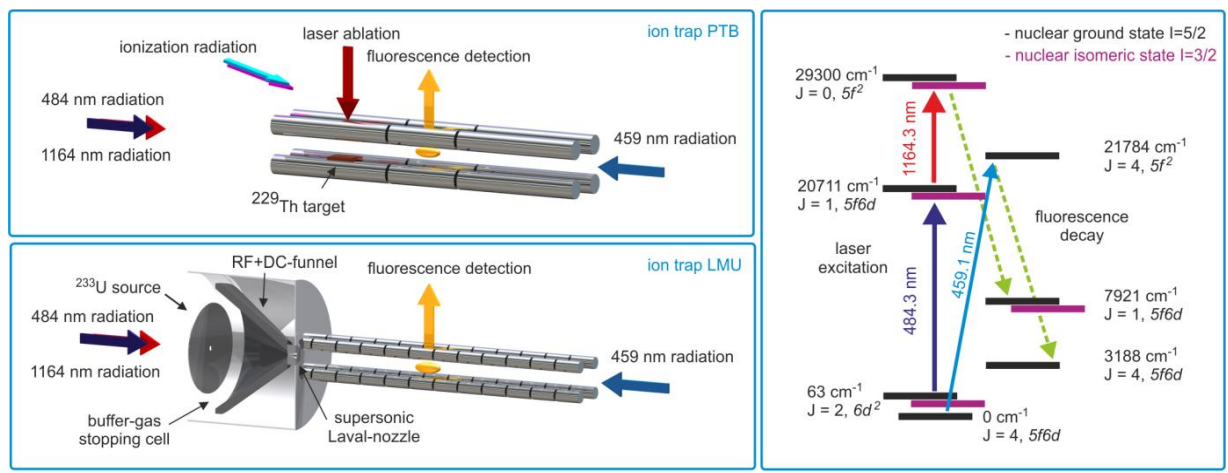

Figure 1: **Experimental setup and $^{229}$Th$^{2+}$ level scheme.** The left side shows schematically the configurations of ion sources, ion traps and laser beams. The transitions and electronic configurations of Th$^{2+}$ levels relevant to the experiment are depicted on the right side, labelled by their energy in cm$^{-1}$ and the electronic angular momentum $J$, which are the same for both nuclear states (see Methods for details). Solid arrows indicate laser excitation and dashed lines show fluorescence decay channels.

## 1 Laser spectroscopy technique

Among the charge states from Th$^+$ to Th$^{3+}$, which are extracted from the $^{233}$U source, Th$^{2+}$ is selected for the experiments. It is the most suitable because of its high yield from the recoil source, the long lifetime of the isomer[25], and its convenient electronic level structure which allows for hyperfine spectroscopy with diode lasers with background-free fluorescence detection in the visible spectral range.

For high-resolution spectroscopy of the hyperfine structure of $^{229}$Th$^{2+}$ we use two independent linear radio-frequency (rf) ion traps[25,27] (see Fig. 1 and Methods). One of the traps (located at Physikalisch-Technische Bundesanstalt, PTB, Braunschweig) is loaded with Th$^+$ by laser ablation from a target containing $^{229}$Th and $^{232}$Th. Three-photon ionization of trapped Th$^+$ is used to produce Th$^{2+}$. The second trap (located at Ludwig-Maximilians-Universität, LMU, München) is loaded with $^{229}$Th recoil ions from the α decay of $^{233}$U, where the isomeric state is populated via a 2% decay branch[9] (for details of the generation of the $^{229m}$Th$^{q+}$ ion beam see Methods). Therefore, the trapped ion cloud consists of a mixture of ions in the ground and the isomeric nuclear states. Daughter products of the $^{233}$U decay chain are also trapped, but do not disturb the spectroscopic measurement (see Methods for details). The combination of the measurements in both traps allows us to identify clearly the hyperfine components of $^{229m}$Th, appearing only with the trapped recoil ions, and to measure isotope and isomer shifts. In both traps, the ions are cooled close to room temperature by collisions with a high-purity buffer gas (argon at PTB, helium at LMU, see Methods). Recording an excitation spectrum by scanning a laser across the Doppler-broadened lines yields a resolution of about 700 MHz. This does not allow to resolve the HFS of $^{229}$Th$^{2+}$ lines and to distinguish resonances of $^{229m}$Th$^{2+}$ from those of $^{229}$Th$^{2+}$. As an example, the lower part of Fig. 2 shows a line of $^{232}$Th ($I = 0$, no HFS splitting) and an unresolved HFS lineshape of $^{229}$Th. For higher resolution we use two-step laser excitation, free from Doppler-broadening[28,29]. The first laser excites a narrow velocity class of ions out of the thermal distribution to an intermediate state, where they are probed by resonant excitation to a higher-lying level using a second tunable laser. Sensitive fluorescence detection of these ions is performed using decay channels at other wavelengths, free from background of laser stray light.

For the two-step laser excitation we choose the transition from the $63_2$ electronic state to $29300_0$ via the $20711_1$ intermediate state (the states are labelled by their energy in cm$^{-1}$ and the electronic angular momentum $J$ as subscript, as shown in Fig. 1). The population of the $63_2$ state is in equilibrium with the electronic ground state ($0_4$) in a ratio of approximately 0.4 through collisions with the buffer gas. The excitation scheme using levels with $J = 2 \rightarrow 1 \rightarrow 0$ consists of a small number of HFS components and therefore simplifies the analysis of the isomeric HFS: while nine hyperfine resonances are present for the nuclear ground state ($I = 5/2$), the isomer with nuclear spin $I = 3/2$ (ref. 7) is expected to appear with eight hyperfine resonances (see Extended Data Fig. 1).

The spectroscopy part of the experimental setup consists of three external-cavity diode lasers (ECDL). The two-step excitation is provided by ECDLs at 484 nm and 1164 nm. Their beams are overlapped and aligned along the trap axis. The third diode laser at 459 nm is used for single-photon excitation of Th$^{2+}$ from the $0_4$ ground state to the $21784_4$ state to monitor the amount of Th$^{2+}$ ions in the trap in order to normalize the fluorescence signals observed from the different HFS components. This is required because the ion number decreases in time due to chemical reactions and charge exchange with impurities in the buffer gas. The wavelengths of the lasers at 484 nm and 459 nm are stabilized to a Fizeau wavemeter, which is periodically calibrated by a diode laser that is stabilized to a resonance line of $^{85}$Rb at 780 nm. To measure the frequency detuning of the 1164 nm laser during the scanning, a temperature stabilized confocal cavity placed in vacuum is used. The fluorescence detection is provided by photomultipliers operating in photon counting mode.

The lower part of Fig. 2 shows the single-photon excitation spectrum of the transition from $63_2$ to $20711_1$ obtained by scanning the frequency of the ECDL at 484 nm over the resonances of $^{232}$Th and $^{229}$Th (see Methods). This data allow us to measure the isotopic shift and to determine the search range of the isomer HFS. The spectrum in the upper part of Fig. 2 shows individual HFS components of $^{229}$Th$^{2+}$ obtained in the second excitation step by scanning the ECDL at 1164 nm over the $20711_1 \rightarrow 29300_0$ transition. For a systematic search for the unknown frequencies of the isomer resonances and for the quantitative analysis of the HFS spectra a mapping of the two-step resonances is performed. The frequency of the 484 nm laser is tuned within the Doppler-broadened HFS of the $63_2 \rightarrow 20711_1$ line in 35 discrete steps of $\approx$ 120 MHz. For each frequency step, the second laser at 1164 nm is scanned continuously over a range $\geq 4$ GHz to provide the detection of the two-step HFS resonances. The frequency of the first step laser is stabilized at each position of the map to the Fizeau wavemeter with a resulting absolute instability of $\leq 5$ MHz. The full width at half maximum (FWHM) of the two-step resonances obtained in the PTB trap is 70 MHz using argon as a buffer gas and 40 MHz at LMU using helium. The two-step excitation mapping is recorded twice, with co- and counterpropagating laser beams in order to substantiate the identification of the HFS components. Because the expected isomer signal is in the range of only 2% of the signal from ions in the nuclear ground state, we choose averaging times of typically four hours per spectrum.

**2 Detection of the isomeric HFS**

In the PTB trap we detect all nine resonances of the nuclear ground state hyperfine structure. A typical record of the two-step nuclear ground state HFS resonances for copropagating beams is shown in Fig. 3 with red points. The second signal (blue points) in the same figure shows the HFS signal obtained in the LMU trap, where a small fraction of the ions is in the isomeric state. The HFS resonances of the isomeric state are clearly observed in comparison with the data acquired at PTB. Fig. 4 shows four resonances of the isomeric state HFS in a logarithmic scale for a different frequency of the first step laser at 484 nm.

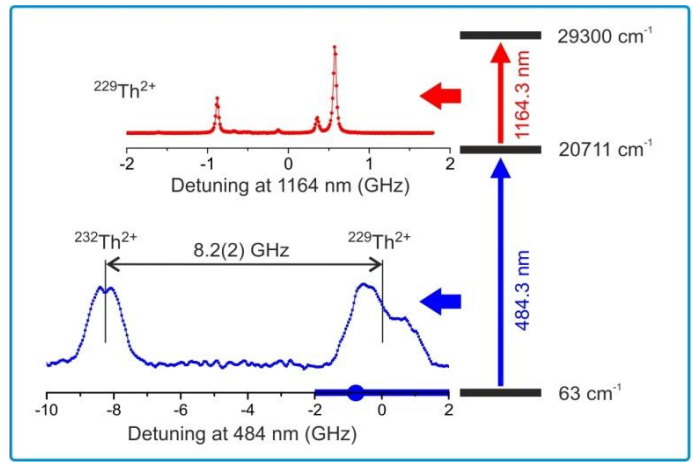

Figure 2. **Two-step excitation spectra.** The Doppler-broadened fluorescence signal of the first excitation step (in blue) obtained with single-photon spectroscopy shows the isotopic shift of 8.2(2) GHz between $^{232}$Th$^{2+}$ and $^{229}$Th$^{2+}$. The upper spectrum (in red) is one exemplary measurement obtained by frequency scanning of the second step laser, when the frequency of the first laser is fixed at -800 MHz detuning (indicated by the blue dot) with respect to the $^{229}$Th HFS centre. Measurements are performed for 35 discrete steps of the laser at 484 nm, indicated schematically by the blue bar on the frequency axis.

In total, we observe seven out of eight resonances of $^{229m}$Th$^{2+}$ in both, co- and counter-propagating beam configurations (see Extended Data Fig. 3). The 8$^{th}$ resonance amplitude is calculated to be small with respect to the signal-to-noise ratio achieved in the experiment. The fraction of the ions in the isomeric state is obtained via the proportion of integrated fluorescence signals of isomeric and nuclear ground state resonances and is determined to be 2.1(5) % (see Methods and Extended Data Fig. 4). This confirms the previously assumed branching to the isomeric state, which was inferred from γ spectroscopy.

We observe a collision-induced change of intermediate state HFS population due to interactions with the buffer gases, appearing for example as resonances c1 and c2 in Fig. 3. The magnitude of this effect is significantly higher with He in the LMU trap than with Ar in the PTB trap. The positions of these resonances can be calculated based on the measured hyperfine splitting of the intermediate state, enabling us to identify them in the spectra. This prevents a wrong assignment of resonances originating from collisions as isomeric HFS. The width of these resonances is approximately 1.5 times larger than the width of the nine main HFS resonances. Furthermore, the amplitude ratio between these resonances and the direct two-step resonances significantly drops with the reduction of the He buffer gas pressure, indicating that these resonances are indeed caused by collisions (see Extended Data Fig. 5).

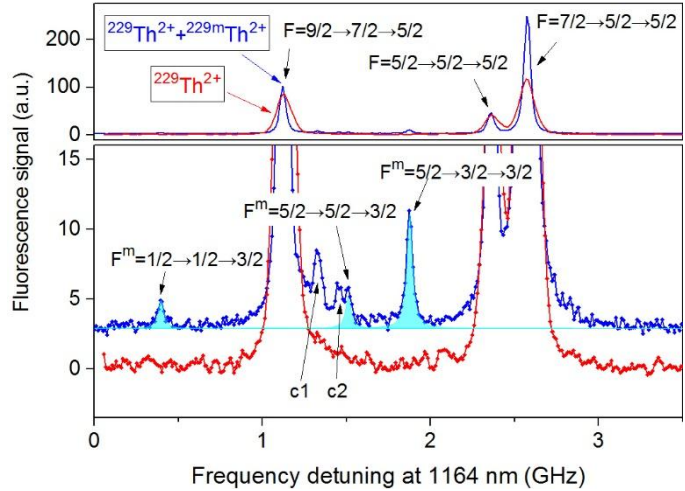

Figure 3. **Comparison of excitation spectra in two traps.** Two-step excitation resonances of Th$^{2+}$, where the first laser is stabilized at -800 MHz detuning with respect to the $^{229}$Th HFS centre and the second laser is scanned. Red and blue colours indicate the signals in the PTB and the LMU trap, respectively. 2 % of the ions in the LMU trap are in the isomeric state. The lower graph shows an enlarged view, with the LMU signal up-shifted for better inspection. Ground and isomeric resonances are shown with the total momenta $F$ and $F^m$ (Extended Data Fig. 1) and the isomeric state is marked in cyan. The resonances c1 and c2 belong to the nuclear ground state and arise from collision-induced change of the intermediate HFS state. The total momenta of the involved states are $F$= 7/2 → 5/2; 7/2 → 5/2 and $F$= 7/2 → 5/2; 3/2 → 5/2, respectively.

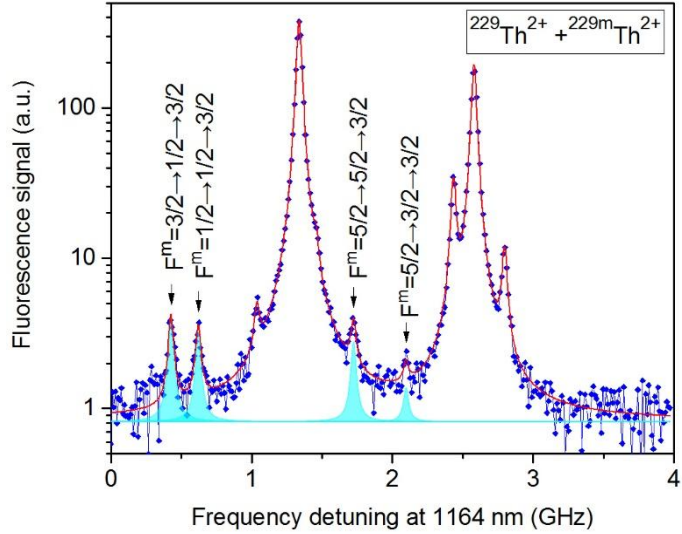

Figure 4. **HFS Resonances of nuclear isomeric and ground states.** Two-step excitation resonances of the nuclear isomer HFS are displayed (marked in cyan colour), showing the relative strengths (in logarithmic scale) and frequency range of nuclear isomeric and ground state resonances (Extended Data Fig. 1). The first laser is stabilized at -260 MHz detuning with respect to the $^{229}$Th HFS centre and the second laser is scanned. The unlabelled peaks belong to the nuclear ground state.

### 3 Isomer properties

The observation of the isomeric HFS allows us to determine the magnetic dipole and quadrupole moments, and the nuclear charge radius. To derive these properties we determine the hyperfine constants $A$ (magnetic dipole) and $B$ (electric quadrupole) of the electronic structure for both ground and isomeric nuclear states. We measure the frequency intervals between the resonances relative to the transmission peaks of the reference cavity and fit them in a least squares adjustment to determine the constants for the $63_2$ and the $20711_1$ electronic states (see Methods). The upper electronic state of the two-step excitation has $J = 0$ and therefore no hyperfine splitting. The results are shown in Table 1. The systematic uncertainty constitutes about 50 % of the total uncertainty of the measurements. Main contributions are caused by an uncertainty of the reference cavity length, instability of the frequency of the first excitation step laser, and the nonlinearity of the second step laser frequency tuning.

Table 1: **Hyperfine constants of $^{229}$Th$^{2+}$ and $^{229m}$Th$^{2+}$ for the electronic levels $63_2$ and $20711_1$.**

| Level [cm$^{-1}$] | Nuclear ground state | | Nuclear isomeric state | |
|---|---|---|---|---|
| | A [MHz] | B [MHz] | A$^m$ [MHz] | B$^m$ [MHz] |
| 63 | 151(8) | 73(27) | -263(29) | 53(65) |
| 20711 | 88(4) | 897(14) | -151(22) | 498(15) |

The measured ratio $A^m/A = -1.73(25)$ determines the magnetic dipole moment $\mu^m$ of the isomer according to the relation $\mu^m = \mu A^m I^m/(AI)$, where $\mu$ indicates the magnetic moment of the nuclear ground state and $I^m$, $I$ are the spin values of the isomeric and nuclear ground states, respectively (see, for example ref. 30). Here we neglect the hyperfine structure anomalies[31,32] which are small for $d^2$, $fd$, and $f^2$ electronic configurations of $Th^{2+}$ levels (see Fig. 1). Therefore, the ratio of the magnetic moments for the isomeric and nuclear ground states is $\mu^m/\mu = -1.04(15)$. The nuclear magnetic moment $\mu$ of the ground state has been measured in two experiments[33,34], with the most precise value $\mu = 0.360(7)\mu_N$ resulting from high-precision calculations[15] and measurements of $^{229}Th^{3+}$ HFS[34]. Based on this value we derive the magnetic dipole moment of the isomeric state as $\mu^m = -0.37(6)\mu_N$. An estimate based on the leading Nilsson configuration had predicted[18,35] $\mu^m = -0.076\mu_N$. The discrepancy indicates that the simplifying Nilsson approach is insufficient for quantitatively characterizing the isomer, since it neglects for example the expected collective quadrupole-octupole coupling of the nuclear deformation (see also refs 20,22).

The spectroscopic quadrupole moment of the isomeric state $Q_s^m$ is determined by $Q_s^m = Q_s B^m/B$, where $Q_s$ is the spectroscopic quadrupole moment of the nuclear ground state. We use only the constants obtained for the $20711_1$ electronic state to derive $B^m/B$ (see Table 1). The measured ratio of the spectroscopic quadrupole moments is $Q_s^m/Q_s = 0.555(19)$. $Q_s$ has been measured in two independent experiments to be $3.15(3)$ $eb$ (ref. 12) and $3.11(6)$ $eb$ (refs 15,34). Using the weighted mean value of these measurements, the quadrupole moment of the isomer is $Q_s^m = 1.74(6)$ $eb$. The spectroscopic quadrupole moment is related to the intrinsic quadrupole moment $Q_0$ through[30] $Q_s = Q_0(3K^2 - I(I+1))/((I+1)(2I+3))$, resulting in $Q_0^m = 8.7(3)$ $eb$ for the isomeric state and $Q_0 = 8.8(1)$ $eb$ for the nuclear ground state. Both states are the bandheads of their rotational bands and therefore the projection of the nuclear spin on the symmetry axis $K$ is equal to $I$. To within the uncertainty, the intrinsic quadrupole moments of both states are the same ($Q_0^m/Q_0 = 0.99(4)$). Therefore, the nuclear charge distribution has a similar prolate shape in both states. This is in good agreement with theoretical predictions[21,22].

To investigate the change of the charge radius between the nuclear ground state and the isomeric nucleus, the isomeric shifts of the first and second excitation steps are derived from the centres of the hyperfine structures, calculated by setting $A=B=0$. The isomeric HFS is shifted to lower frequency relative to $^{229}Th^{2+}$ by $0.29(3)$ GHz for the electronic transition $63_2 \rightarrow 20711_1$. The isotope shift of this line between $^{229}Th^{2+}$ and $^{232}Th^{2+}$ isotopes is $8.2(2)$ GHz (see Fig. 2). The analysis for the second excitation step $20711_1 \rightarrow 29300_0$ yields the isomeric shift of $0.21(5)$ GHz and the isotopic shift of $6.2(3)$ GHz. The average ratio of isomer and isotope shifts for both transitions is $0.035(4)$. The isotopic shift for the heavy ions is primarily determined by the field shift with only small corrections arising from the mass shift[36] and therefore, is directly related to the nuclear charge radius. The measured isotope shifts correspond to the difference in the mean-square charge radii $\langle r^2 \rangle^{232} - \langle r^2 \rangle^{229} = 0.33(5)$ fm$^2$ (ref. 29), with $\langle r^2 \rangle^{229} \approx (5.76\text{ fm})^2$. Consequently, the difference in the mean-square radii of the isomeric and nuclear ground state in $^{229}Th$ is $\langle r^2 \rangle^{229m} - \langle r^2 \rangle^{229} = 0.012(2)$ fm$^2$.

**4 Discussion**

The unique transition between nuclear ground and isomeric states in $^{229}Th$ creates a bridge between atomic and nuclear physics and opens new perspectives for fundamental physics as well as for technology progress in metrology. We have measured the $^{229m}Th$ nuclear moments and isomeric shifts of two $^{229m}Th^{2+}$ lines. The determination of these basic nuclear properties allows one to calculate the hyperfine structure of $^{229m}Th$ for any electronic transition when the frequencies of $^{229}Th$ and another isotope with measured nuclear radius[29] are known. For the future, this makes it possible to apply the sensitive electronic-nuclear double resonance detection[4] of the isomer in the search for nuclear laser excitation and in nuclear clock operation with trapped ions. This is particularly important when the long radiative lifetime of the isomer makes it difficult to detect photons emitted in the isomer decay, and constitutes an important step towards the development of a nuclear clock.

With the provided quantitative information on the nuclear moments, analyses of expected systematic uncertainties of $^{229}$Th nuclear clocks can be stated more precisely[4,5,37,38]. In a clock based on $^{229}$Th-doped crystals[37,38] the nuclear quadrupole moment interacts with crystal field gradients, which may lead to a significant frequency shift. In the trapped ion clock the field-induced shifts will only make minor contributions to the uncertainty budget. The $^{229}$Th nuclear clock has been proposed as a particularly sensitive system to search for temporal variations of the fine structure constant α (ref. 39). This is based on a model where the small nuclear transition energy of $E_{is} \approx 8$ eV appears as the result of the nearly perfect cancellation of a change in the Coulomb energy $\Delta E_C = E_C^m - E_C \approx -1$ MeV with opposite and nearly equal changes of the nuclear energy from the strong interaction. Such a cancellation would be very sensitive to the values of the coupling constants of the electromagnetic and strong forces. The model has been criticised, based on the reasoning that the transition is performed by an unpaired neutron, leaving the Coulomb energies of $^{229}$Th and $^{229m}$Th essentially equal[40]. Following Ref. 41, the change in Coulomb energy can be expressed in terms of quantities that have been measured here: $\Delta E_C = -485$ MeV $((\langle r^2 \rangle^{229m}/\langle r^2 \rangle^{229})-1) + 11.6$ MeV $((Q_0^m/Q_0)-1) = -0.29(43)$ MeV (see Methods). The uncertainty in $\Delta E_C$ is dominated by the contribution from the ±4% uncertainty in $Q_0^m/Q_0$. Although this result is not sufficient to prove $|\Delta E_C| \gg E_{is}$, it appears likely that the α−sensitivity of the $^{229}$Th nuclear clock exceeds those of present atomic clocks by several orders of magnitude. Together with the expected high accuracy of the nuclear clock, this will provide a chance to test predictions for temporal variations of coupling constants and to experimentally assess theories unifying gravity with other interactions.

**Acknowledgments**. We thank C. Mokry, J. Runke, K. Eberhardt, N. G. Trautmann for the production of the $^{233}$U source and M. Ehlers, S. Hennig, K. Kossert for the PTB $^{229}$Th source. We acknowledge discussions with Chr. Tamm, B. Lipphardt and thank T. Leder, M. Menzel, A. Hoppmann for providing expert technical support. We acknowledge financial support from the European Union's Horizon 2020 Research and Innovation Programme under Grant Agreement No. 664732 (nuClock), from DFG through CRC 1227 (DQ-mat, project B04) and TH956-3-2 and from the LMU Department of Medical Physics via the Maier-Leibnitz Laboratory.

**Author Contributions.** J.T., M.V.O., developed the spectroscopy lasers; J.T., P.G., M.V.O., L.v.d.W., B.S., D.M.M., P.G.T. did preparatory experimental work and performed the spectroscopy experiment; J.T., P.G., M.V.O., and E.P. provided the data analysis; M.V.O., P.G.T. and E.P. supervised the experiment; the $^{233}$U source was produced in the group of C.E.D. All authors discussed the results. M.V.O., J.T., P.G., and E.P. wrote the manuscript with input from all authors.

**Methods**

**Ion trapping.**

**a) PTB trap.** The radio-frequency linear Paul trap located at PTB[27] is loaded with ≈ $10^5$ Th$^+$ ions from a Th(NO$_3$)$_4$ solution (containing $^{229}$Th and $^{232}$Th in about equal proportions) dried on a tungsten target, by laser ablation using a Nd:YAG laser emitting 5 ns pulses with an energy of ≤ 1 mJ at 1064 nm. Doubly charged thorium ions are generated in the trap via three-photon resonant ionization of Th$^+$. For the first ionization step a 402 nm external cavity diode laser (ECDL) with an output power of 20 mW, shaped into pulses of 90 ns length by an acousto-optical modulator, is used to pump the 24874 cm$^{-1}$ Th$^+$ state. The second and third ionization step photons are provided by the third harmonic generation (THG) of a pulsed

Ti:Sa laser (pulse length of 20 ns, THG peak power of $\approx$ 1 kW) via the 63258 cm$^{-1}$ Th$^+$ state[42]. Both lasers operate at a repetition rate of 1 kHz and pulses of the Ti:Sa and the ECDL are overlapped in time. The continuous resonant ionization produces a cloud of $\approx 10^3$ $^{229}$Th$^{2+}$ ions and compensates the loss of $^{229}$Th$^{2+}$ ions due to formation of molecules with impurities of the buffer gas. Using argon as buffer gas at 0.1 Pa pressure, we cool ions to room temperature and depopulate metastable states by collisional quenching. The experiments are performed using Ar because of a higher trap loading efficiency achieved in comparison with He as a buffer gas. For the photodissociation of Th$^+$ compounds formed in the trap (see ref. 42) we use the 4$^{th}$ harmonic radiation of a Q-switched Nd:YAG laser at 266 nm with a pulse energy of $\approx$ 10 μJ (for the scheme of the optical setup see Extended Data Fig. 2). Due to the presence of $^{229}$Th and $^{232}$Th isotopes in the solution on the target we trap both isotopes simultaneously.

**b) LMU trap.** The radio frequency quadrupole ion trap located at LMU is loaded with α-recoil thorium ions from a $^{233}$U source (source 3 in ref. 25). In the α decay of $^{233}$U, the low-energy isomeric state in $^{229}$Th is populated via an estimated 2% decay branch. Therefore, the trapped ion cloud consists of a mixture of ions in the ground and isomeric nuclear states. The $^{233}$U source was produced at the Institute for Nuclear Chemistry of the University of Mainz by molecular plating[43,44]. 290 kBq of $^{233}$U were deposited with 90 mm diameter onto a Ti-sputtered Si-wafer of 100 mm outer diameter and 0.5 mm thickness. The $^{233}$U material contains a number fraction of below $10^{-6}$ of $^{232}$U and was chemically purified by ion-exchange chromatography before deposition (25 month prior to the experiment) in order to remove the $^{233}$U and $^{232}$U daughter isotopes. To allow for the laser-spectroscopy experiments, the $^{233}$U source possesses a central hole of 8 mm diameter[45].

About $10^5$ $^{229}$Th α-recoil ions are leaving the $^{233}$U source per second, due to their kinetic energy of about 84 keV. They are stopped in a buffer-gas stopping cell filled with $3.2\cdot 10^3$ Pa of ultra pure He 6.0, which is further purified by catalytic purification and a cryotrap[45,46]. During the stopping process, charge exchange occurs, producing predominantly thorium ions in the 2+ and 3+ charge states. After thermalization, these ions (together with the α decay daughter products of the $^{233}$U decay chain) are guided by an electric RF+DC-funnel system, consisting of 50 ring electrodes, towards a de-Laval extraction nozzle with a 0.6 mm diameter nozzle throat. The extraction nozzle forms a supersonic gas jet and drags the ions into the subsequent (12-fold segmented) RF quadrupole, operated in this experiment as an ion trap. Due to an extraction efficiency of $\sim$ 5% and $\sim$ 10% in the 2+ and 3+ charge states, respectively, more than $10^4$ $^{229}$Th ions are entering the RF quadrupole per second. In this way the trap is continuously loaded with $^{229}$Th ions, however only about $10^3$ $^{229}$Th$^{2+}$ ions are actually trapped due to the trap's limited loading capacity. Besides $^{229}$Th, also other isotopes originating from α decay of nuclides other than $^{233}$U (e.g. from $^{232}$U and its decay chain) or from sputtering can potentially enter the ion trap. These are listed in Extended Data Tab. 2, together with an estimation of their relative abundance compared to $^{229}$Th. Most of them do not play a role as a potential source of background, as their relative abundance is too small. Isotopes that could affect the experiment are discussed in the sections 'Exclusion of $^{230}$Th isotope' and 'Exclusion of coincident absorption lines'. Due to charge capture, the $^{229}$Th$^{3+}$ ions are reduced to $^{229}$Th$^{2+}$ after a few seconds. The He pressure in the trapping region is reduced to about 0.1 Pa by a differential pumping stage. To avoid the cumulation of molecular compounds in the trap formed due to chemical reactions with impurities, we periodically empty the RF trap every 75 s. A new detection cycle starts 15 s after the trap reset, when the amount of $^{229}$Th$^{2+}$ ions in the trap reaches its maximum value of about $10^3$ ions.

**Spectroscopic lasers.** The spectroscopy of the thorium isomer hyperfine structure is provided by external cavity continuous wave diode lasers with a typical linewidth of 100 kHz. The ECDL at 1164 nm has an output power of 30 mW and a tuning range of about $\geq$ 4 GHz. The blue ECDLs at 459 nm and 484 nm provide an output power of ≈ 15 mW and a tuning range of ≥ 15 GHz. The radiation of the lasers is delivered to the trap by single mode polarization maintaining fibres. The power of all lasers in both traps is about 4 mW due to losses in fibre coupling and clipping by the supersonic nozzle, corresponding to an intensity of 1.5 W/cm² for each beam. The scheme of the optical setup is shown in the Extended Data Fig. 2.

**Laser frequency stabilization.** To avoid long term frequency drifts and to provide controlled frequency steps of the 484 nm laser with an accuracy of a few MHz its radiation is stabilized to a Fizeau wavemeter (HighFinesse WS7) by a computer-based locking system. The Rb-stabilized ECDL at 780 nm is used to calibrate the wavemeter in intervals of 1000 s. The 780 nm ECDL is stabilized to the $^2S_{1/2}$ $F = 3 \rightarrow$ $^2P_{3/2}$ $F = 4$ $^{85}$Rb line by the modulation transfer spectroscopy technique (see, for example ref. 47).

**Fluorescence detection.** The fluorescence of the excited Th$^{2+}$ ions is detected using a photomultiplier tube (PMT) with a bandpass interference filter that transmits in the range of 445 ± 23 nm, which corresponds to one of the decay channels of the upper state and provides a background-free detection for the two-step excitation. A second PMT is used for the measurements of the isotopic shift or alternatively to provide measurements of the amount of thorium ions in the trap during the two-step excitation. It is set to either detect the decay channels of the excitations at 484 nm or at 459 nm, using a filter at 643±10 nm or 540±8 nm, respectively, which block the laser stray light. Counters are used to register the PMT signals. For the PTB trap the photon counting is terminated during the ionization and dissociation laser pulses to prevent their influence on the spectroscopic signal.

**Hyperfine structure.** If a nucleus has a spin $I > 1/2$, it may possess a magnetic dipole moment and an electric quadrupole moment. The interaction of the valence electrons with these moments cause the hyperfine splitting of the electronic levels, where the energy shift of the individual level is determined by $E_{HFS}$ $(JIF) = AK/2 + B[(3/4)K(K+1)-I(I+1)J(J+1)] / [2I(2I-1)J(2J-1)]$, with $K = F(F+1) - J(J+1) - I(I+1)$. Here, $J$ is the electronic angular momentum and $F$ the total momentum. The hyperfine constants $A$ and $B$ are determined by the magnetic dipole and electric quadrupole interactions (see, for example ref. 30, 48). For $^{229}$Th, the nuclear ground state has the spin $I = 5/2$ and the isomer $I = 3/2$. This leads to a splitting of the 63$_2$ electronic level into five sub levels for the nuclear ground state and four sub levels for the isomeric state. The 20711$_1$ level consists of three hyperfine levels in both nuclear states. Following the selection rules for electric dipole transitions ($\Delta F = 0, \pm 1$), the two-step excitation spectrum with $J = 2 \rightarrow 1 \rightarrow 0$ for the nuclear ground and isomeric states consists of nine and eight resonances, respectively (see Extended Data Fig. 1).

**Two-step excitation resonances.** Assume laser 1 has a frequency detuning $\Delta v_1$ with respect to the centre of the first step HFS. $\Delta v_{gi}$ and $\Delta v_{ie}$ are the frequency shifts of the individual hyperfine components of the first ($gi$) and the second ($ie$) step from the centres of their HFS. The velocity class ($v$) of particles of the individual HFS component, which is excited to the intermediate state, is described as $k_1v = -\Delta v_{gi} + \Delta v_1$. Particles with the same $v$ are excited to the upper state according to the equation $\pm(k_2v) = -\Delta v_{ie} + \Delta v_2$, where $\Delta v_2$ is the second laser frequency detuning from the second step HFS centre, $+k_2v$ corresponds to co- and $-k_2v$ to counterpropagating beams. Here $k_{1,2}$ are the wavevectors of the first and the second excitation steps. Combining both excitation steps for ions with the same $v$, the position of an individual narrow two-step resonance (free from Doppler-broadening) is represented as $\Delta v_2 = \pm k_2/k_1 [-\Delta v_{gi} + \Delta v_1] + \Delta v_{ie}$. The amplitude of the resonance depends on the fraction of ions within the velocity group of the Doppler distribution, which interacts with the first step laser radiation and the product of $\Omega_{2gi}$ and $\Omega_{2ie}$, where $\Omega$ are Rabi frequencies of the transitions of both excitation steps (see, for example, ref. 28).

**HFS constants.** The intervals between the resonances are measured with respect to the transmission peaks of the stable, confocal cavity, which are recorded simultaneously with the HFS resonances. The measured frequency intervals of the HFS are fitted in a least-squares adjustment according to the equations described in the sections above to determine the *A* and *B* constants for the $63_2$ and the $20711_1$ states. The algorithm attempts fits for all viable assignment combinations of *F* values for the resolved transitions. The fitting for the isomeric state is provided without imposing fixed proportions $A_m/A$ and $B_m/B$ which have to occur for both electronic states.

**Map of the HFS resonances.** Extended Data Fig. 3 shows selected records demonstrating the evolution of the HFS peaks of the ground and isomeric states for different frequency positions of the first excitation step. The isomeric state HFS resonances are marked with red colour labels and the description of all resonances is given in Extended Data Table 1. The evaluation of the positions and the amplitudes of the newly observed seven peaks over the mapping indicates that those resonances fit to the HFS pattern of a $I = 3/2$ isomer. We performed tests for the appearance of spurious resonances from the collision-induced intermediate state changes, possible laser multimode operation, and back reflection of the spectroscopic laser beams (mixing of co- and counterpropagating beam geometries). These tests confirm the observed resonances to be original spectroscopic features.

The Extended Data Fig. 4 is generated by tracking the positions and amplitudes of the second excitation step resonances for all acquired spectra of both nuclear states. The figure shows the relative amplitudes of the resonances and is used to calculate the fraction of ions in the isomeric state.

**Exclusion of $^{230}$Th isotope.** Estimations for daughter products from the $^{233}$U source (Extended Data Table 2) show upper limits for the presence of $^{228}$Th and $^{230}$Th isotopes. For all transitions the resonance of $^{230}$Th ($I = 0$, no HFS) should appear between the lines of $^{232}$Th and $^{229}$Th, like those of the $^{229m}$Th HFS. The isotopic shift of $^{230}$Th$^{2+}$ with respect to $^{229}$Th$^{2+}$ for the 484 nm trasition is 3.2 GHz, calculated from previous measurements at 459 nm using the LMU trap with a $^{234}$U source and from ref. 29. This is outside the range of the observed isomeric HFS. In the experiment with the $^{233}$U source the $^{230}$Th$^{2+}$ resonance is not observed, indicating a significantly smaller flux than listed in Extended Data Table 2.

**Exclusion of coincident absorption lines.** Because the $^{233}$U source emits a variety of ions of different elements (as α decay daughter products originating in the $^{233}$U decay chain), which are loaded into the trap simultaneously with thorium ions (See Extended Data Table 2), spectral lines of those elements might be detected in parallel to the isomer signal. There are only two elements (U, Pu) which have an extracted flux high enough to possibly be detected. For the first excitation step the nearest uranium and plutonium lines originating from low lying levels are detuned by $\approx$ 200 GHz ($^{238}$U$^+$, $\tilde{v}$=20640.51 cm$^{-1}$, transition from $915_{9/2}$ to $21555_{9/2}$) and $\approx$ 60 GHz ($^{240}$Pu$^+$, $\tilde{v}$=20645.62 cm$^{-1}$, transition from $3970_{5/2}$ to $24615_{3/2}$) (refs 49–51). We exclude also the influence of Th$^{2+}$He complexes, potentially formed by interaction with the buffer gas, due to the estimated low binding energy of less than 7500 cm$^{-1}$ (see, for example ref. 52), which would be dissociated by laser excitation at 484 nm (20648 cm$^{-1}$). Moreover, since the signal we detect requires coincidence of two resonance conditions, we can exclude that the recorded thorium spectra are affected by other elements.

**Isomeric lifetime in Th$^{2+}$.** It is in principle possible to determine the isomeric lifetime in Th$^{2+}$ by measuring the time evolution of the amplitudes of the resonances for both nuclear states. The isomeric signal will show an additional exponential decay due to its finite lifetime. This experiment is so far limited by the storage time of the ions in the trap of $\approx$ 60 s, determined by chemical reactions and charge exchange with impurities in the buffer gas. Therefore, this value can only be given as a lower limit for the isomer lifetime. By improving the storage time by two orders of magnitude, measuring the isomeric lifetime becomes feasible, providing an important parameter of the clock.

**Sensitivity to the value of the fine structure constant.** The $^{229}$Th nuclear clock has been proposed as a particularly sensitive system to search for temporal variations of the fine structure constant $\alpha$ (ref. 39), but the proposal has been met with scepticism[40]. The sensitivity $(\alpha/f)\,df/d\alpha$ of the nuclear transition frequency to the value of $\alpha$ is equal to the ratio $\Delta E_C/E_{is}$ of the change in Coulomb energy between the ground and isomeric state, and the total transition energy ($\approx$ 7.8 eV) between ground state and isomer. Theoretical estimations of $\Delta E_C$ from nuclear structure calculations[22,40] vary from a few keV to a few MeV. It has been proposed to calculate $\Delta E_C$ via the change of the nuclear charge radius and electric quadrupole moment from measured isomer shift and HFS data[41]. Treating the nucleus as a uniform, hard-edged, prolate ellipsoid, one obtains $\Delta E_C$ = -485 MeV $((\langle r^2 \rangle^{229m} / \langle r^2 \rangle^{229})-1)$ + 11.6 MeV $((Q_0^m/Q_0)-1)$ based on eq. (7) of Ref. 41, using updated values of $Q_0$ and $\langle r^2 \rangle^{229}$. Applied to our data, this results in $\Delta E_C$ = -0.29(43) MeV. Since the change of the charge radius $\Delta \langle r^2 \rangle / \langle r^2 \rangle \approx 4 \cdot 10^{-4}$ is small, the uncertainty in the change of the quadrupole moment, now known from $Q^m_0/Q_0$ = 0.99(4), yields the dominant uncertainty contribution to $\Delta E_C$. While smaller values cannot be excluded with certainty, the most likely modulus for the $\alpha$-sensitivity of a $^{229}$Th nuclear clock is about $4 \cdot 10^4$.

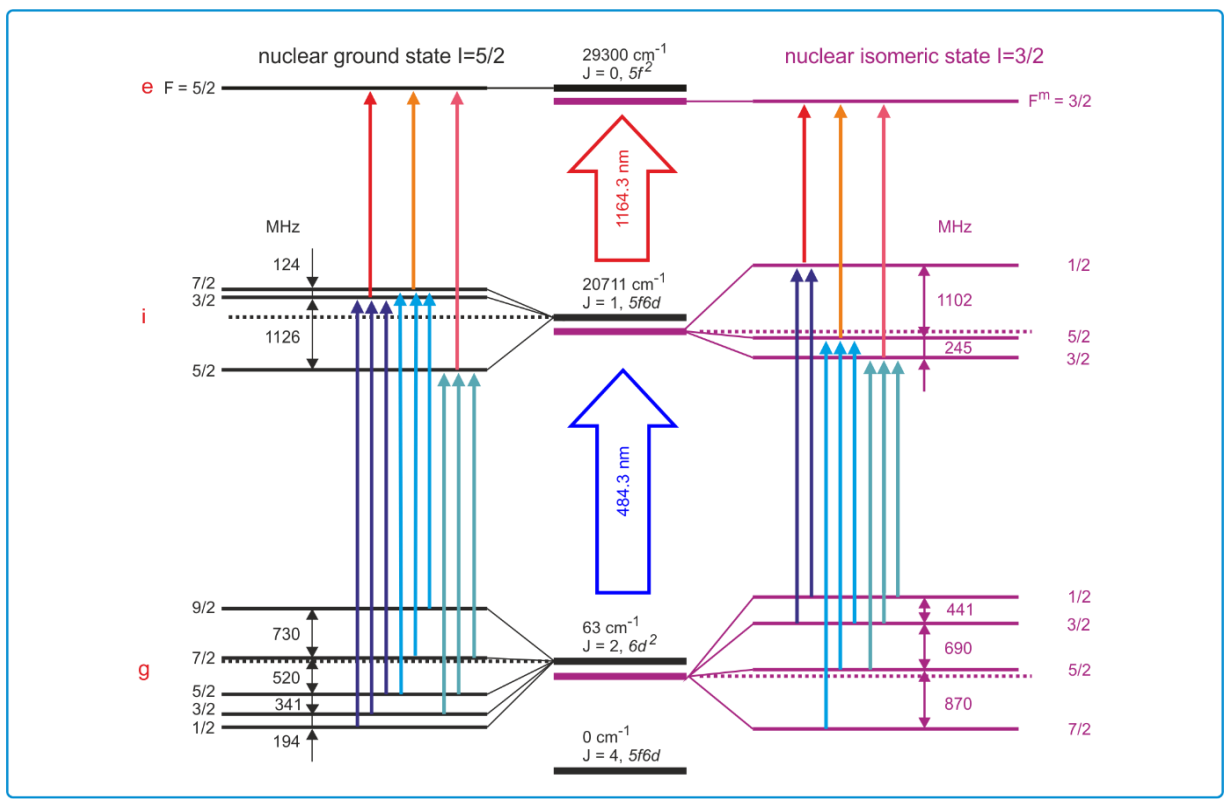

Extended Data Figure 1. **Detailed level scheme of the two-step excitation.** Transitions and electronic configurations of initial (g), intermediate (i), and excited (e) states relevant to the experiment, labelled by their energy in cm$^{-1}$ and the electronic angular momentum *J*. Hyperfine sub levels are indicated by their total angular momentum *F* and *F$^m$*. Transitions belonging to the same intermediate hyperfine level are depicted with the same colour. The hyperfine intervals are calculated from the *A* and *B* hyperfine constants presented in Table 1.

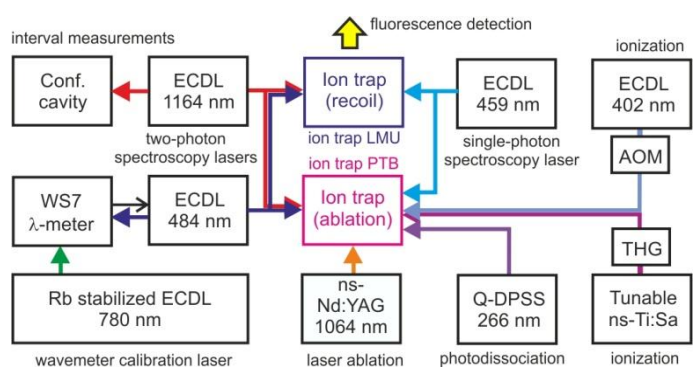

Extended Data Figure 2. **Scheme of the optical setup.** The spectroscopy laser of the first step excitation (484 nm) is locked to the wavemeter, which is calibrated by a Rb-stabilized ECDL at 780 nm. The second step laser (1164 nm) tuning is monitoring with the confocal cavity. The ECDL at 459 nm is used to detect the amount of ions in the traps. The loading of Th$^{2+}$ in the PTB trap is provided by ablation (ns-Nd:YAG 1064 nm) and further three-photon ionization (1$^{st}$ step: 402 nm ECDL, 2$^{nd}$/3$^{rd}$ steps: THG of ns-Ti:Sa).

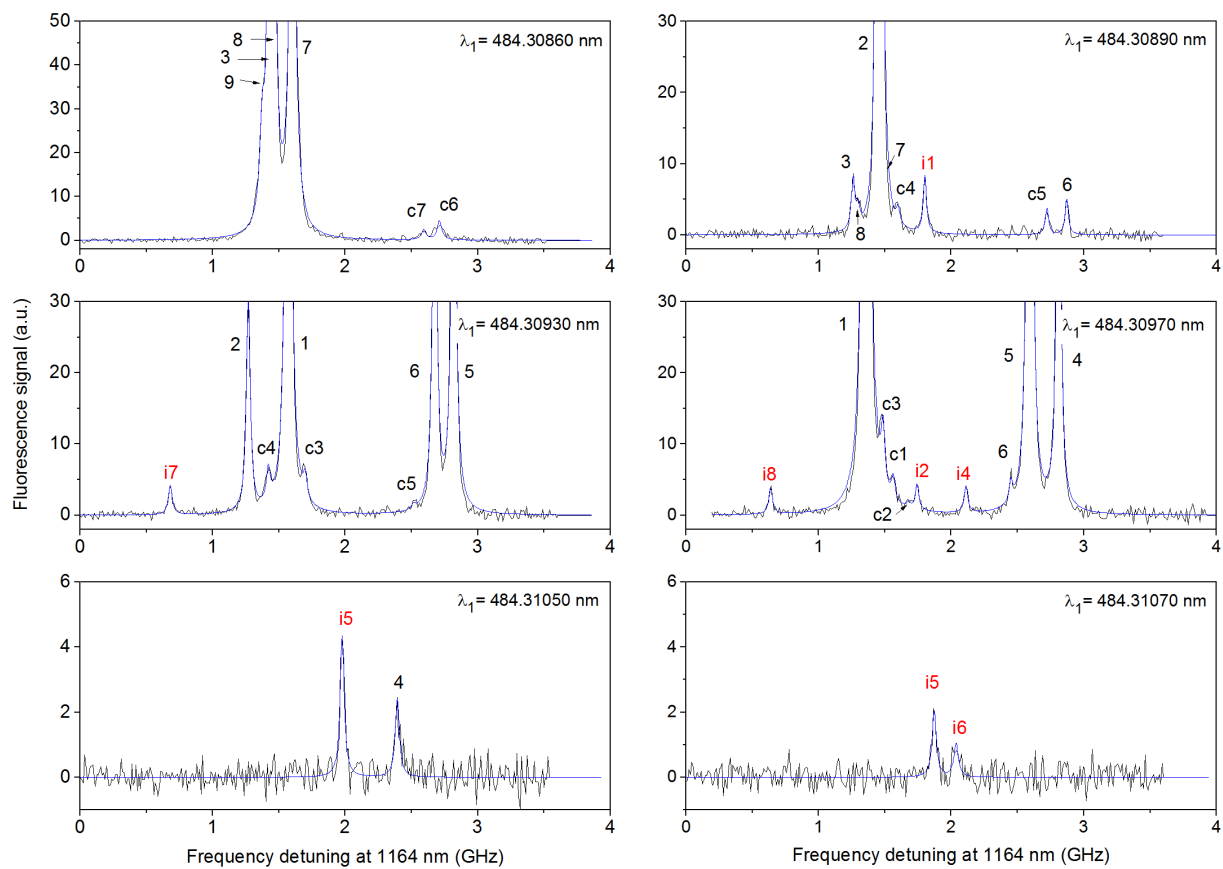

Extended Data Figure 3. **Selected spectra of two-step excitation.** The records of the resonances for different positions of the ECDL at 484 nm showing the observed isomeric peaks for the copropagating case (labelled "i"). The resonances which originate from collisions of ions in the intermediate state are labelled "c". The description of the peaks with their total angular momenta is given in the Extended Data Table 1.

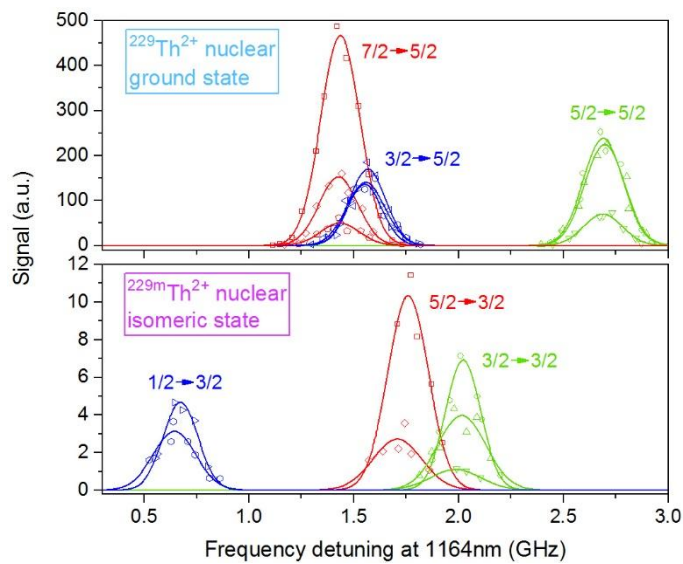

Extended Data Figure 4. **Mapping of the second excitation step.** The experimental points present the amplitudes and the positions of the two-step resonances obtained setting the laser at 484 nm on a certain frequency and tuning the laser at 1164 nm. The frequency of the 484 nm laser is changed in ≈ 120 MHz steps. The groups of the resonances which are shown with the same color correspond to the transitions from the same intermediate state with the total angular momentum *F*, populated from different ground state hyperfine components. The upper and the lower graphs demonstrate the HFS transitions of $^{229}$Th$^{2+}$ in the nuclear ground and the isomeric states, respectively.

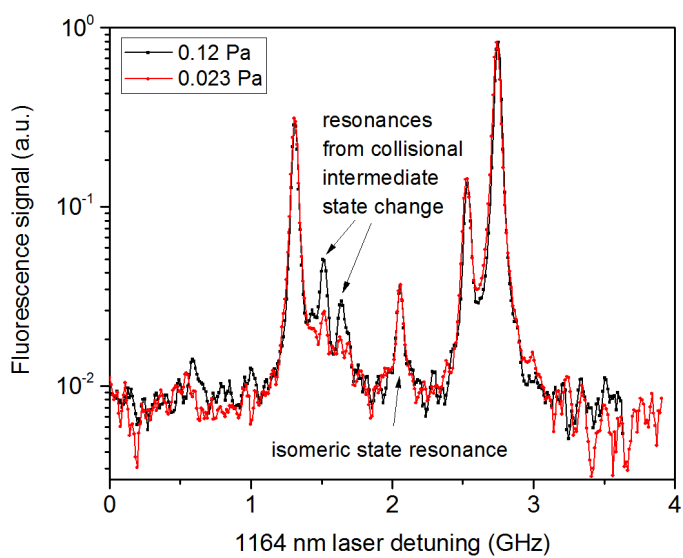

Extended Data Figure 5. **Pressure dependence of collision-induced intermediate HFS state changes.** Two-step excitation resonances of Th$^{2+}$, where the first laser is stabilized at -800 MHz detuning with respect to the $^{229}$Th HFS centre and the second laser is scanned. The measurement is performed for two different He buffer gas pressures, showing a decrease in relative amplitude of the collisional resonances for the reduction of the buffer gas pressure. Note that the isomeric resonance is not affected by the change in He pressure.

Extended Data Table 1. **Systematics of the observed resonances.** The detected resonances are listed with the total angular momenta of the electronic states involved in the excitation. The resonances of the nuclear ground state are labelled with numbers. The resonances which arise from collisional changes of the intermediate state population are described with both states involved in the collisions and labelled "c". Isomeric resonances are marked with "i". The resonance i3 (marked with an asterisk) is not observed in the experiment.

| label | $F_g \to F_i \to F_e$ | label | $F_g \to F_i$ ; $F'_i \to F_e$ | label | $F^m_g \to F^m_i \to F^m_e$ |
|---|---|---|---|---|---|
| 1 | $9/2 \to 7/2 \to 5/2$ | c1 | $7/2 \to 5/2$; $7/2 \to 5/2$ | i1 | $7/2 \to 5/2 \to 3/2$ |
| 2 | $7/2 \to 7/2 \to 5/2$ | c2 | $7/2 \to 5/2$; $3/2 \to 5/2$ | i2 | $5/2 \to 5/2 \to 3/2$ |
| 3 | $5/2 \to 7/2 \to 5/2$ | c3 | $9/2 \to 7/2$; $3/2 \to 5/2$& | i3* | $3/2 \to 5/2 \to 3/2$ |
| 4 | $7/2 \to 5/2 \to 5/2$ |    | $5/2 \to 5/2$; $3/2 \to 5/2$ | i4 | $5/2 \to 3/2 \to 3/2$ |
| 5 | $5/2 \to 5/2 \to 5/2$ | c4 | $7/2 \to 7/2$; $3/2 \to 5/2$ | i5 | $3/2 \to 3/2 \to 3/2$ |
| 6 | $3/2 \to 5/2 \to 5/2$ | c5 | $7/2 \to 7/2$; $5/2 \to 5/2$ | i6 | $1/2 \to 3/2 \to 3/2$ |
| 7 | $5/2 \to 3/2 \to 5/2$ | c6 | $5/2 \to 3/2$; $5/2 \to 5/2$ | i7 | $3/2 \to 1/2 \to 3/2$ |
| 8 | $3/2 \to 3/2 \to 5/2$ | c7 | $3/2 \to 3/2$; $5/2 \to 5/2$ | i8 | $1/2 \to 1/2 \to 3/2$ |
| 9 | $1/2 \to 3/2 \to 5/2$ |    |    |    |    |

Extended Data Table 2. **Extraction of isotopes from the $^{233}$U source.** Amount of recoils leaving the $^{233}$U source by α-recoil and recoil sputtering. Given extraction ratios show the upper limit of recoils relative to the amount of $^{229}$Th recoils.

| Isotope | Extraction (rel. to Th-229) | Isotope | Extraction (rel. to Th-229) |
|---|---|---|---|
| Th-229 | 1 | Pu-238 | $4 \times 10^{-6}$ |
| U-233 | ≈ 1 | Pu-239 | $2 \times 10^{-3}$ |
| Th-229 decay chain | $2 \times 10^{-4}$ | U-235 | $1 \times 10^{-2}$ |
| U-232 | $6 \times 10^{-7}$ | Pu-240 | $4 \times 10^{-4}$ |
| Th-228 | $1 \times 10^{-3}$ | U-236 | $1 \times 10^{-2}$ |
| Th-228 decay chain | $7 \times 10^{-4}$ | Pa-231 | $2 \times 10^{-5}$ |
| U-234 | $2 \times 10^{-2}$ | Ac-227 | $1 \times 10^{-4}$ |
| Th-230 | $7 \times 10^{-3}$ | Ac-227 decay chain | $6 \times 10^{-6}$ |